# In which fields are citations indicators of research quality?[1]

Mike Thelwall, Kayvan Kousha, Emma Stuart, Meiko Makita, Mahshid Abdoli, Paul Wilson, Jonathan Levitt. University of Wolverhampton, UK.

Citation counts are widely used as indicators of research quality to support or replace human peer review and for lists of top cited papers, researchers, and institutions. Nevertheless, the relationship between citations and research quality is poorly evidenced. We report the first large-scale science-wide academic evaluation of the relationship between research quality and citations (field normalised citation counts), correlating them for 87,739 journal articles in 34 field-based UK Units of Assessment (UoAs). The two correlate positively in all academic fields, from very weak (0.1) to strong (0.5), reflecting broadly linear relationships in all fields. We give the first evidence that the correlations are positive even across the arts and humanities. The patterns are similar for the field classification schemes of Scopus and Dimensions.ai, although varying for some individual subjects and therefore more uncertain for these. We also show for the first time that no field has a citation threshold beyond which all articles are excellent quality, so lists of top cited articles are not pure collections of excellence, and neither is any top citation percentile indicator. Thus, whilst appropriately field normalised citations associate positively with research quality in all fields, they never perfectly reflect it, even at high values.

## 1 Introduction

Citations are widely used as formal or informal indicators of research value. In Italy, for example, articles with enough citations and published in a journal with enough citations per paper used to be automatically classified as high quality in the national research assessment exercise (Abramo & D'Angelo, 2016). Citations also inform peer review exercises (REF2021, 2020), organisations such as Clarivate celebrate highly cited articles and researchers, and university league tables often include a citation-based component (Waltman, et al., 2012). Citation-based Journal Impact Factors (JIFs) and variants are frequently used for recognition and reward too (McKiernan et al., 2019). Nevertheless, the value of citations for research assessment is contested and controversial. Research evaluation guidelines caution against overreliance on research metrics (CoARA, 2022; Hicks et al., 2015; Wilsdon et al., 2015), and there are strong arguments against using citations for aspects of research assessment because they do not always reflect impact and ignore some article influences (e.g., MacRoberts & MacRoberts, 2010). Many organisations have also signed the San Francisco Declaration of Research Assessment (DORA), which argues against reliance on journal impact factors (sfdora.org). In the UK, most national research evaluation avoids all bibliometrics as unhelpful to guide evaluations of quality (REF2021, 2020). Thus, despite the widespread use and avoidance of citation counts and their controversial nature, it is still not clear which fields they are appropriate for and how they relate to research quality. This article provides the first peer reviewed science wide article-level evidence of this. It uses a large sample of journal articles with norm referenced expert review quality scores from the UK.

---

Research quality is a vague overall concept but is usually thought of in terms of methodological rigour, novelty/originality, and impact on science or society (Langfeldt et al., 2020). The way in which these three aspects is tested varies substantially between fields. For example, to be judged 'world leading', research might need to be "a primary or essential point of reference" in the arts and humanities or impress in terms of "the scale, challenge and logistical difficulty posed by the research" for health-related studies (REF2021, 2020). Of the three aspects of research quality, citations best reflect impact on science (Aksnes et al., 2019), so citation counts may undervalue research that is particularly strong for rigour, novelty or impact on society. Since there may be a tendency for research to be simultaneously strong or weak in all aspects of quality, and this may vary between fields, it is not clear whether citation counts are reasonable indicators of overall quality in any or all fields.

A core theoretical basis for using citation counts as an indicator of research quality, or at least its scholarly impact dimension, is that citations serve to acknowledge relevant or foundational prior work of other scholars. Thus, counting the citations to an article might give a measure of how often it has proven useful (Merton, 1973). There are several arguments against this, however. First, there are many reasons to cite prior work, including for background context, to refute, and to show improvement without necessarily drawing upon the cited work (Lyu et al., 2021). Second, humanities fields are non-hierarchical and there is much less need to build on the work of other scholars (Whitley, 2000). Third, there are many factors that influence the choice of citations, such as the tendency to cite known scholars, friends, or editors, biasing the counts (Borgman & Furner, 2002; Vinkler, 1987). Fourth, non-journal outputs (e.g., reports, books, art) are important in some fields (Hicks, 2004) but are often largely or fully excluded from citation databases. A statistical response to criticisms like these is to accept that there are reasons for citing that do not reflect impact or that reflect little impact, but that when citations are aggregated on a sufficiently large scale then the "imperfections" may tend to average out. This would allow indicators based on average citations to have some value, even if they do not work well for individual journal articles (van Raan, 2004). Since the amount of bias and the amount of "signal to noise" in citations is unknown, the task of identifying the contexts, such as fields and years, in which it is appropriate to use citation-based indicators is essentially a statistical one: assessing if and when citation counts correlate to a sufficient degree with article quality.

Many studies have compared public aggregate evidence of research quality with average citations for collections of outputs, with mixed results. Rankings of UK departments based on average peer review scores for their outputs have been compared to average citation-based rankings, with correlations being very strong (rho=0.9) for psychology (Smith & Eysenck, 2002) library and information science (rho=0.8) (Norris & Oppenheim, 2003; Oppenheim, 1995), Archaeology (rho=0.7), Genetics (rho=0.7), and Anatomy (rho=0.5) (Seng & Willett, 1995), and Music (rho=0.8) (Oppenheim & Summers, 2008) and political science (partial correlation: 0.5) (Butler & McAllister, 2009). A larger scale study found strong associations between average citations and average REF scores for journal articles in life and health sciences (except nursing), business and economics, but weak associations in the social sciences (Mahdi et al., 2008). High correlations (0.7-0.8) between departmental REF2014 output rankings and median citations per paper have also been found for ten UoAs (Pride & Knoth, 2018).

Outside the UK, an investigation into 12,000 Italian research articles correlated institutional average peer review scores with institutional average citations per paper in ten fields. There were strong correlations in most, including Physics (rho=0.8), Earth Sciences

(0.8), Biology (0.7), and Chemistry (0.6) (Franceschet & Costantini, 2011), but weaker correlations have been found with a different method for Italy (except medicine, 0.5: Abramo et al, 2011; Baccini & De Nicolao, 2016). High correlations have also been obtained for the Netherlands (Rinia et al., 1998; van Raan, 2006). From a related perspective, panel ratings had weak correlations with citation-based indicators for research groups within an institution in Norway (Aksnes & Taxt, 2004). These studies give little information about the strength of article level correlations within fields, however, because correlations increase in magnitude when data are aggregated, with the degree of increase depending on the size of the aggregation units. Thus, it is not possible to draw conclusions about article-level correlations from institution-level correlations.

A few studies have directly investigated the extent to which citation counts correlate with research quality for journal articles. The largest scale study was non-academic (not peer reviewed, written by two professional statisticians) investigated peer review scores for about 25,000 journal articles published in 2008 with citation-based indicators in 36 UK Research Excellence Framework (REF) Units of Assessment (UoAs) and reported weaker results for articles from 2013 (HEFCE, 2015). Overall, REF peer review scores for individual articles significantly and positively correlated (0.3) with Elsevier's field normalised citation impact metric Source-Normalised Impact per Paper (SNIP), Field-Weighted Citation Impact (0.3), and citation counts (0.2). There were large disciplinary differences within this overall figure, with the strongest correlations between citation counts and REF scores in Clinical Medicine (rho=0.7), Chemistry (0.6), Physics (0.6) and Biological Sciences (0.6). Correlations in most social sciences, arts and humanities were typically below 0.3 and some were negative but unreliable due to small sample sizes (e.g., 15 articles for one correlation) (HEFCE, 2015). Because of its goals, this study included duplicate articles (the same article submitted by authors in different institutions), which undermines the general (i.e., non-REF) value of the correlations because multiply-submitted articles can expect to be both higher quality and more cited because they have more authors (including in the UK: Thelwall & Maflahi, 2020). Also the reliance on REF self-classifications for articles is imperfect because multidisciplinary authors and department members with out-of-field specialisms (e.g., medical statisticians) could result in articles submitted to inappropriate UoAs. The arts and humanities data also included a minority of articles due to a majority of missing DOIs. Nevertheless, this is probably the best available evidence of the relationship between research quality and citation counts at a relatively fine-grained level, but the methods had the problems reported above and the results did not report confidence intervals and only used a single year of data, limiting the conclusions that can be drawn. Moreover, UK UoAs are unique aggregation units that do not easily map onto other field classification schemes, which also limits the generalisability of the results.

Given the lack of science-wide academic evidence about the relationship between research quality and citations in the different academic fields, this article addresses the following research questions.
- RQ1: In which fields do more cited standard journal articles (excluding reviews) tend to be higher quality? In other words, in which fields is there a positive correlation between citation scores and quality scores.
- RQ2: Does the answer to the above depend on the field classification scheme used?
- RQ3: Is there a citation score threshold in any field, above which all research is high quality? Here, "high quality" is equated with the REF "world leading" definitions, as discussed below. The answer to this is relevant to lists of top cited articles and attempts to use citation-based thresholds to identify excellent research.
- RQ4: What is the overall shape of the relationship between citations and research quality? In other words, what shapes exist in graphs of quality scores against citation rates? This is important because non-monotonic shapes suggest that the overall relationship between citations and research quality differs from the pattern for different citation ranges.

## 2 Peer review in the Research Excellence Framework 2021

The UK REF2021 can claim to be the largest scale, most expensive and most financially important science wide academic peer review exercise ever conducted in the world. The 34 disciplinary subpanels of the REF assessed 185,594 outputs (mainly journal articles) from 76,132 academic staff organised into 1876 submissions (each roughly a university department) as well as 6,781 impact case studies and information about the scholarly environments of 157 UK higher education institutions (REF2021, 2022a). The administrative cost of REF2014 was estimated to be £240 million (Technopolis, 2015). REF scores direct 2 billion pounds in research funding per year (UKRI, 2022ab), so 14 billion pounds in total for the seven years for which each REF's scores are active.

From initial planning to eventual publication of results, each REF takes at least eight years. For example, the REF2021 results were published on the 8[th] of May, 2022 but one of the early public planning exercises was the 2014-15 Independent Review of the Role of Metrics in Research Assessment and Management that produced the Metric Tide report (Wilsdon et al., 2015), and the current paper is an offshoot from a study commissioned in 2021 preparing for REF2028. The care with which the REF is designed can be seen from the 13 public background documents that informed the transition from REF2014 (REF2021, 2018), with the most influential being the Stern report that recommended, amongst other things, that each research active scholar should submit 1-5 outputs for assessment (rather than submission of researchers being optional, but with 4 outputs each in REF2014, excluding double-counting outputs). The UK higher education sector is extensively consulted on any proposals for REF changes, with 388 responses to the Stern report alone (REF2021, 2018).

At the heart of REF2021 is the scoring of the 185,594 outputs by over 1000 experts organised into 34 Units of Assessment (UoAs) from UoA 1 Clinical Medicine to UoA 34 Communication, Cultural and Media Studies, Library and Information Management. These experts are nominated by institutions following a public call for specific expertise areas (REF2021, 2021a; REF2021, 2021b). The experts are trained in systems, ethics, and assessment procedures and their working methods are outlined in a 106-page public document. This includes overall and panel-specific definitions of quality and their applicability to the four level scoring criteria used (REF2021, 2020). Each output is initially allocated by subpanel (UoA) chairs to two experts who independently score it, then consult and agree on a score on a nine-point scale, optionally consulting bibliometrics in cases of disagreement in 11 subpanels. These scores are then discussed collectively in each subpanel and there are also main panel calibration discussions combining multiple UoA subpanels, and REF-wide statistical checks on score distributions to norm reference the scores. At some stage the nine-point scale is narrowed down to the 4 point scale (plus 0 for out of scope) that is eventually published. The extensive norm referencing is essential to the credibility of the system and is useful for bibliometric uses of the data since it allows interdisciplinary analyses. Unfortunately, the individual output scores had to be deleted in 2022 for legal reasons but the provisional data was temporarily released for research to the project that produced this article, the only time this has been allowed for REF data.

As a result of the above process, each of the 185,594 REF outputs were allocated a quality score for "originality, significance and rigour" of 1* "recognised nationally", 2* "recognised internationally", 3* "internationally excellent", or 4* "world-leading". Outputs judged ineligible or below national quality were scored 0 instead (REF2021, 2019). In addition to this overall REF definition/interpretation of quality, there are more specific criteria for each of the four Main Panels, each of which contains multiple UoAs (REF2021, 2020). For example, the criteria below apply to the mainly health and life sciences UoAs 1 to 6 in Main Panel A:

> The sub-panels will look for evidence of some of the following types of characteristics of quality, as appropriate to each of the starred quality levels: • scientific rigour and excellence, with regard to design, method, execution and analysis • significant addition to knowledge and to the conceptual framework of the field • actual significance of the research • the scale, challenge and logistical difficulty posed by the research • the logical coherence of argument • contribution to theory-building • significance of work to advance knowledge, skills, understanding and scholarship in theory, practice, education, management and/or policy • applicability and significance to the relevant service users and research users • potential applicability for policy in, for example, health, healthcare, public health, food security, animal health or welfare. (REF2021, 2020)

The remaining three main panels have specific criteria for each of the starred levels. For example, the highest quality (i.e., 4*) Main Panel C (UoAs 13 to 24, mainly social sciences) guidance is:

> In assessing work as being four star (quality that is world-leading in terms of originality, significance and rigour), sub-panels will expect to see some of the following characteristics: • outstandingly novel in developing concepts, paradigms, techniques or outcomes • a primary or essential point of reference • a formative influence on the intellectual agenda • application of exceptionally rigorous research design and techniques of investigation and analysis • generation of an exceptionally significant data set or research resource. (REF2021, 2020)

In contrast, the lowest (i.e., 1*) grade for Main Panel C equates to the following:

> In assessing work as being one star (quality that is recognised nationally in terms of originality, significance and rigour), sub-panels will expect to see some of the following characteristics: • providing useful knowledge, but unlikely to have more than a minor influence • an identifiable contribution to understanding, but largely framed by existing paradigms or traditions of enquiry • competent application of appropriate research design and techniques of investigation and analysis. (REF2021, 2020)

Despite the assessor expertise, the detailed guidelines and repeated norm-referencing, the REF2021 output scores are imperfect. The main reason is that the 1000+ experts will have substantial topic knowledge gaps, with none having the expertise to assess some of the 185,594 outputs. For example, none of the assessors for UoA 34, which incorporates library and information science, was a bibliometrician. In addition, there may be institutional, gender or other biases in scores, or simple prejudices against competing research paradigms, topics, or methods. Another problem is that each output had two assessors, giving a workload of about 370 outputs to score per assessor, over about a year. This is a substantial task for busy academics, and this seems to preclude a detailed assessment of each output. On the other hand, the articles have already passed journal peer review, so the REF assessors can expect to be primarily reading polished, high quality research.

## 3 Methods

### 3.1 Data

The data analysed in this paper is a subset of the journal articles submitted to UK REF2021. For this, as mentioned above, each active higher education researcher in the UK had to submit between 1 and 5 outputs first published between 2014 and 2020, with an average of 2.5 outputs per full time equivalent member of staff. These outputs were submitted to one of 34 Units of Assessment (UoAs) and were then individually evaluated by at least two UoA subject specialists and awarded one of four quality scores for "originality, significance and rigour". All types of research output could be submitted but only journal articles are analysed here. Review-type outputs are ineligible for the REF, so all articles report original research. Each author of a paper is entitled to submit it, but two authors from the same institution are usually not allowed to submit the same output.

Provisional REF2021 scores were supplied in March 2022 for 148,977 journal articles, which is an almost complete set except those from the University of Wolverhampton for confidentiality (because the authors were from the University of Wolverhampton). There were 34% 4*, 50% 3*, 5% 2*, 1% 1*, and 0.2% 0. The range of scores for each UoA are available online (Figure 3.2.2 of the main report: Thelwall et al., 2022) alongside other background information about the dataset. The articles were spread reasonably evenly between 2014 and 2020, from 11% in 2014 to 16% in 2018. The articles were matched with journal articles in Scopus with a recorded date between 2014 to 2018 that were downloaded in January 2021, to coincide with the date when the REF2021 evaluation was scheduled to start (although it was delayed by Covid-19). The matching was by DOI (99%) or by title, year and journal manually checked (1%). Articles from after 2018 were excluded to give a citation window of at least two years for analysis, and the 318 articles scoring 0 were removed because these had often not been evaluated for quality.

The citation counts for the journal articles were transformed into field and year normalised scores to allow different fields and years to be merged. To calculate Normalised Log-transformed Citation Scores (NLCS) (Thelwall, 2017), all Scopus articles 2014-18 were first log-transformed with ln(1+x) to reduce skewing and prevent the normalisation calculations from being heavily influenced by individual highly cited articles. After this, in each of the 326 Scopus narrow fields and years, the average log-transformed citation count was calculated. Finally, the log-transformed citation count ln(1+x) of each matching REF2021 article was divided by the average just calculated for the field and year in which it was published. Thus, if the citation counts of n articles in a single field are $c_1, c_2, \ldots c_n$ then the NLCS of the $k$th article would be as follows (worked examples are available online: Day 1, Talk 3, slide 20 of http://cybermetrics.wlv.ac.uk/SummerSchoolSeptember2020.html):

$$\ln(1 + c_k) / \sum_{i=1}^{n} \ln(1 + c_i)$$

Articles in multiple fields were divided instead by the average of the averages of the fields containing them. An NLCS for an article of 1 always equates to world average citation count for its Scopus-indexed field(s) and year. Values higher than 1 always mean more cited than average for the publishing field(s) and year. The NLCS values were grouped into fields for analysis and compared with the provisional REF2021 quality scores for the same articles.

REF2021 organises the evaluation in 34 Units of Assessment grouped into four Main Panels, but there are other ways of grouping research into fields and so two alternative categorisation schemes were also used: The article-based Artificial Intelligence (AI) scheme of Dimensions.ai and the mainly journal-based scheme of Scopus. For Dimensions, each REF2021 article with a DOI was matched against Dimensions records with an Application Programming Interface (API) DOI search. The top-level Field Of Research (FOR) codes reported by Dimensions for each matching article were saved for the matching record. For Scopus, the 27 top-level broad fields were used, as recorded by the Scopus API.

After all data processing, there were 87,739 journal articles across the 34 UoAs and 83,327 across the 4 main panels. The reason for the difference is that duplicate articles were eliminated within groups so that each UoA or Main Panel dataset includes no duplicates. There were many duplicates between UoAs (i.e., the same article submitted to multiple UoAs by different authors), so more duplicates were removed when forming main panel groups than UoA groups. For Scopus, a total of 144,207 articles were analysed and for Dimensions 99,661 articles were analysed. In both cases, the figures include duplicates between but not within fields or UoAs.

## 3.2 Analysis

The REF and NLCS scores were compared primarily through Spearman correlations. Although Pearson correlations would have been reasonable, given the log transformation to reduce skewing, Spearman is a conservative option and is appropriate given that REF scores occur on a non-scalar four-point qualitative system. Confidence intervals for the correlations were calculated using standard Fisher (1915) transformations.

For RQ3 and RQ4, the articles were bucketed into groups of at least 25 for analysis. This had two purposes. First, the scores for individual REF outputs are confidential and so individual values cannot be shown. Second, a reasonable sample size is necessary to differentiate between coincidence and trend. For example, to address RQ3, if the most cited article in UoA 1 had a 4* quality score, it would not be reasonable to give a positive answer based on the citation count of that article because 39% of UoA 1 articles score 4* irrespective of citation counts (Figure 3.2.2. of: Thelwall et al., 2022). The bucket size of 25 seems like a reasonable compromise between too fine grained and too broad. This relatively arbitrary bucket size is a limitation, however. There is not a perfect way to select an appropriate sample size without a prior belief about the expected rate of 4* amongst the highest cited articles. For example, if 90% of highly cited articles (however defined) in a UoA were believed to be 4*, then a bucket size of 25 still gives a probability of $0.9^{25}=0.07$ that all the selected articles were 4* by coincidence, but if the belief was 80% then the corresponding probability would be $0.8^{25}=0.003$. For additional context, five buckets of size 25 contained only 4* scores across all 34 UoAs, but none of these were buckets of the highest cited 25 of any UoA.

# 4 Results

## 4.1 RQ1, RQ2: overall relationship between citations and research quality

Articles with more citations tend to be higher quality in all fields of science, whether using the REF (Figure 1), Dimensions (Figure 2) or Scopus (Figure 3) classification schemes. Surprisingly, given the prior HEFCE (2015) findings with smaller sample sizes and other issues, there are statistically significant positive correlations (i.e., 95% confidence intervals that do not include 0) even in most arts and humanities fields, including UoA 33: Music, Drama, Dance and Performing Arts (Figure 1), Studies in Creative Arts and Writing (Figure 2) and Arts and Humanities (Figure 3). Moreover, none of the correlations are negative, unlike in the HEFCE (2015) study.

Increasing the citation window 2 from to 5 years by restricting the articles to those from 2014-15 changes the correlations little but increases all the confidence interval widths (the revised figures for this are not included here since they add little information). Wide confidence intervals in Figures 1 and 3 are due to small sample sizes and so the exact values have little relevance.

Comparing classification schemes, there are some patterns. First, the spread of field-based correlation magnitudes (ignoring the Scopus Multidisciplinary class) is substantially higher for the REF UoAs (0.55) than for Dimensions (0.36) and Scopus (0.39). Correlations for REF UoAs vary between 0.02 and 0.57 (Figure 1), for Dimensions they vary between 0.11 and 0.47 (Figure 2), for Scopus they vary between 0.06 and 0.45 (Figure 3). The relatively wide UoA spread of correlations suggests that the REF field classification scheme could be more effectively clustering articles by field, so that topics for which citations are better indicators of quality are less mixed with topics for which citations are worse indicators of quality. Alternative explanations are also possible, however, and the high correlation for the Scopus Multidisciplinary category is presumably due to predominantly high scores and citations for prestigious generalist journals like Nature and Science in comparison to lower scores and fewer citations to less well-known generalist journals.

From the perspective of individual subjects, the classification scheme has little influence on the strength of correlation for some subjects but not others, when they are comparable. For example, for Mathematical Sciences, the correlations have a spread of 0.04: 0.35 (UoAs), 0.32 (Dimensions) and 0.31 (Scopus Mathematics). In contrast for Chemistry the correlations have a four times larger spread of 0.17: 0.57 (UoAs), 0.40 (Dimensions Chemical Sciences) and 0.42 (Scopus). Again, there are alternative plausible explanations, but it is possible that purer categories allow higher correlations by avoiding work for which citations have little relevance to quality (e.g., perhaps chemical engineering for chemistry).

In terms of the overall disciplinary patterns shown, physics, chemistry, biology, and medicine are the areas with the consistently highest correlations under all three schemes, followed by other natural and health sciences. In contrast, arts and humanities topics have the weakest correlations under all three schemes, with social sciences and engineering tending to be between these two groupings.

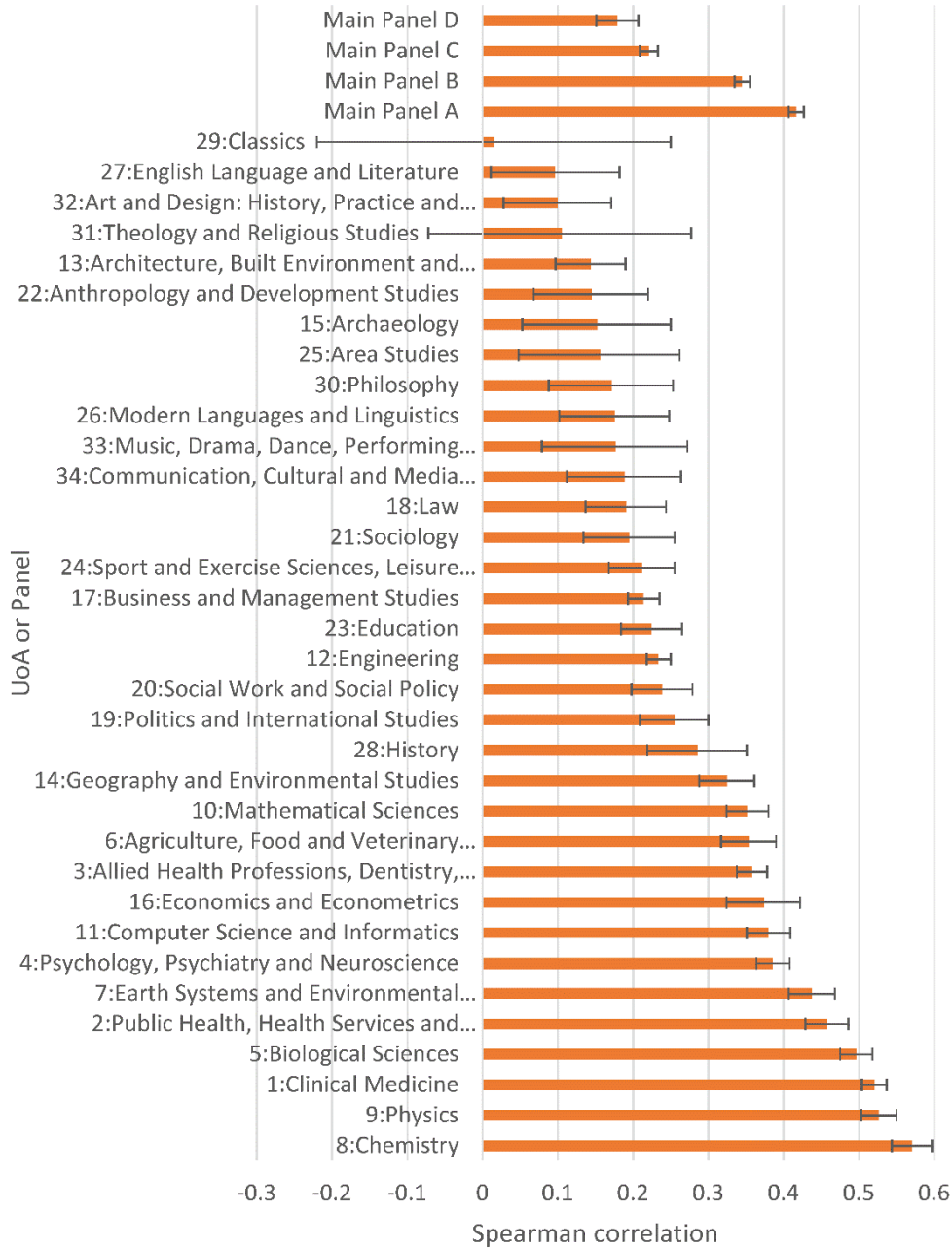

Figure 1. Spearman correlations between field and year normalised citation counts (NLCS) and REF scores for 2014-18 journal articles submitted to UK REF2021 by submitting Unit of Assessment or Main Panel. Error bars indicate 95% confidence intervals.

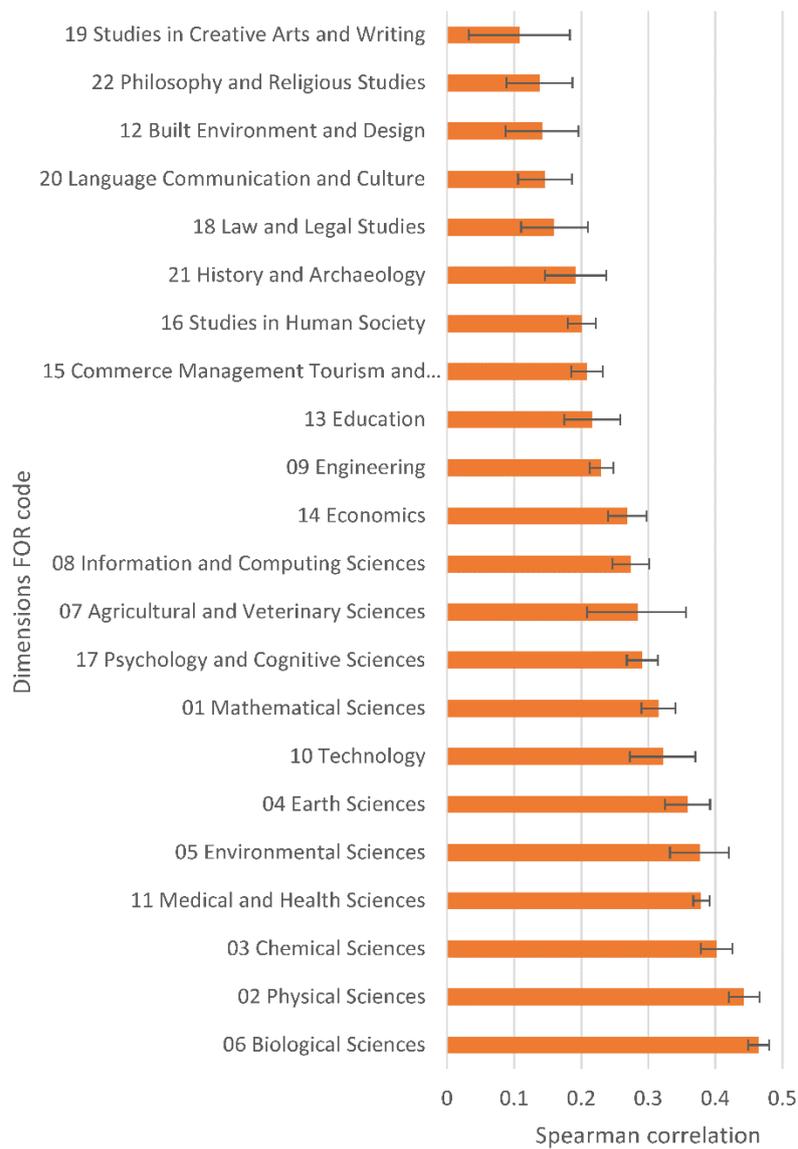

Figure 2. Spearman correlations between field and year normalised citation counts (NLCS) and REF scores for 2014-18 journal articles submitted to UK REF2021 by Dimensions FOR code (n=22). Error bars indicate 95% confidence intervals.

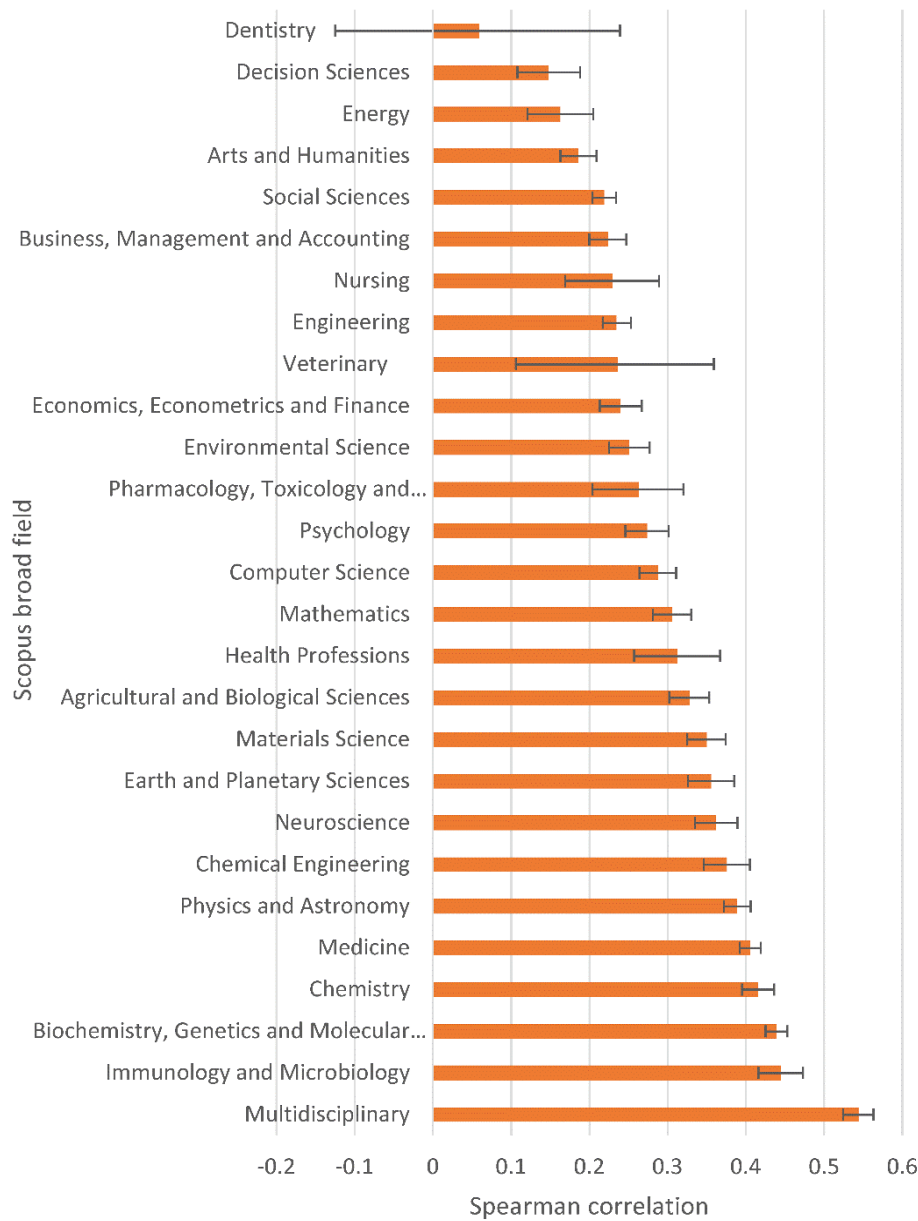

Figure 3. Spearman correlations between field and year normalised citation counts (NLCS) and REF scores for 2014-18 journal articles submitted to UK REF2021 by Scopus broad field (n=27). Error bars indicate 95% confidence intervals.

## 4.2  RQ3: Do high citations guarantee high quality in any field?

In answer to RQ3, after bucketing articles together into groups of size at least 25, there were no UoAs in which the top group all had the highest REF2021 quality score (Figure 4). Thus, the simple answer to the research question is no: there is no citation threshold in any UoA that guarantees the highest quality score, at least for a bucket size of 25. Seven UoAs are close to 100% 4*, however. Increasing the citation window to 3 years (articles from 2014-17, citations from 2021), four years (articles from 2014-16, citations from 2021) or five years (articles from 2014-15, citations from 2021) does not change the result: the most cited (NLCS) 25 articles in each UoA are never always rated 4*.

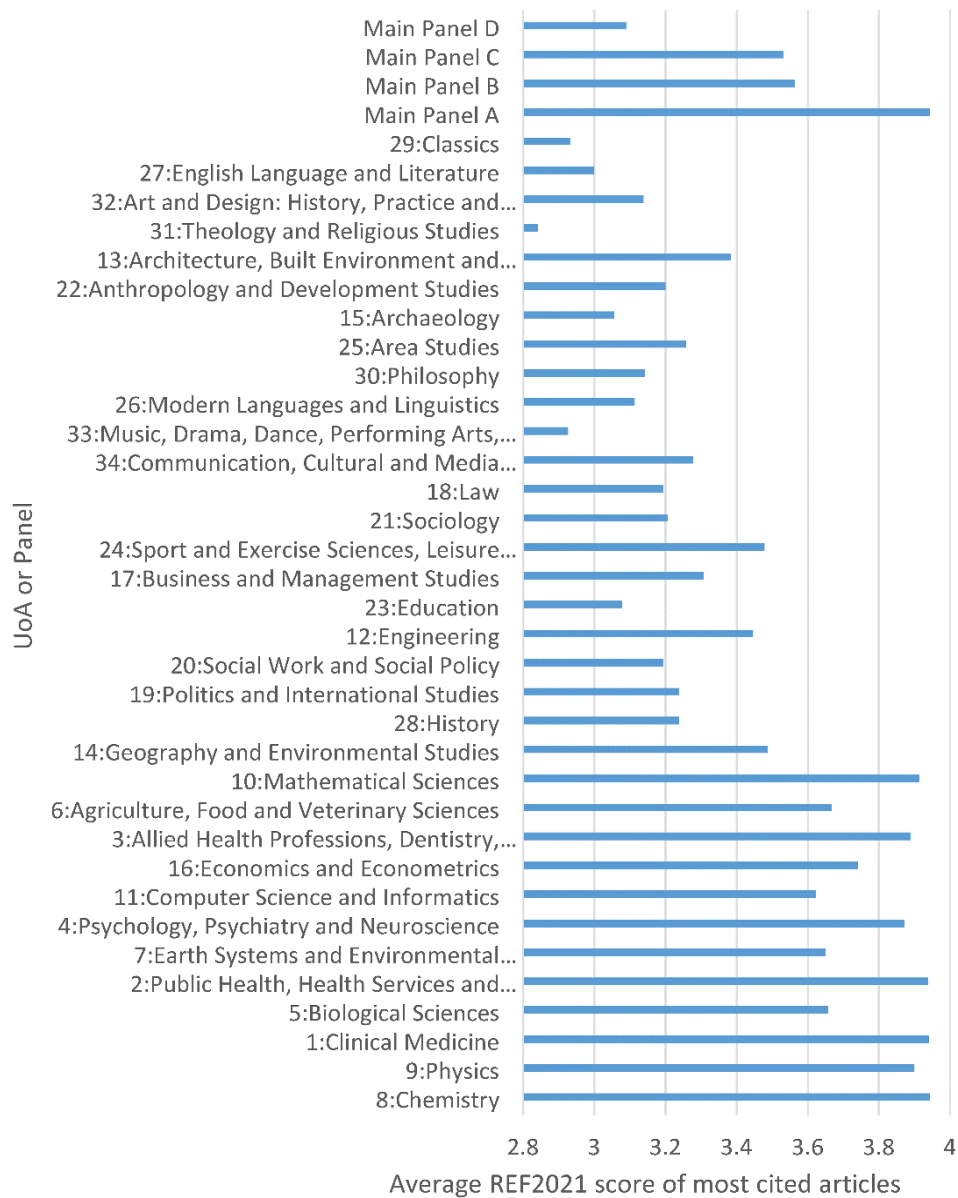

Figure 4. Mean REF scores for 2014-18 journal articles submitted to UK REF2021 for the 25 articles with the highest field and year normalised citation counts (NLCS) by submitting UoA or Main Panel. Sort order as for Figure 1.

### 4.3 RQ4: Overall shape of the relationship between citations and research quality

A positive Spearman correlation can reflect many different underlying shapes, so it is informative to examine the underlying relationship between citations and research quality. The clearest way to do this is to plot average REF2021 scores against NLCS values, bucketing articles into groups with similar NLCS and taking the mean REF2021 score. This hides the variation between articles with similar NLCSs but shows the underlying trend. This is a problematic approach because the scores 1* to 4* are ordered but not scalar. Nevertheless, it is at least plausible to interpret 1* to 4* as a scale 1 to 4 and this assumption is routinely made for departmental Grade Point Averages (GPAs) constructed from REF scores. Given that this aspect of the calculation of GPAs does not seem to be challenged in the UK, it seems reasonable to make the same assumption here.

In all cases where there is a positive correlation above 0.1, the underlying shapes are close to straight lines, but some are more are consistent with approximate logarithmic curves: relatively rapid increase in average REF scores for NLCS increases at lower NLCS values and smaller increases in average REF scores for NLCS increases at higher NLCS values. The steepness of the increase and the range of average REF scores differs substantially between UoAs, however.

In fields with higher correlations (e.g., Figures 5, 6), the increase in average REF score for higher NLCS values is relatively steep, ending close to 4. Although there are variations within each NLCS range, in these fields, citation scores seem to be good indicators of quality and it would be possible but surprising to find an excellent little cited article or a non-excellent highly cited article.

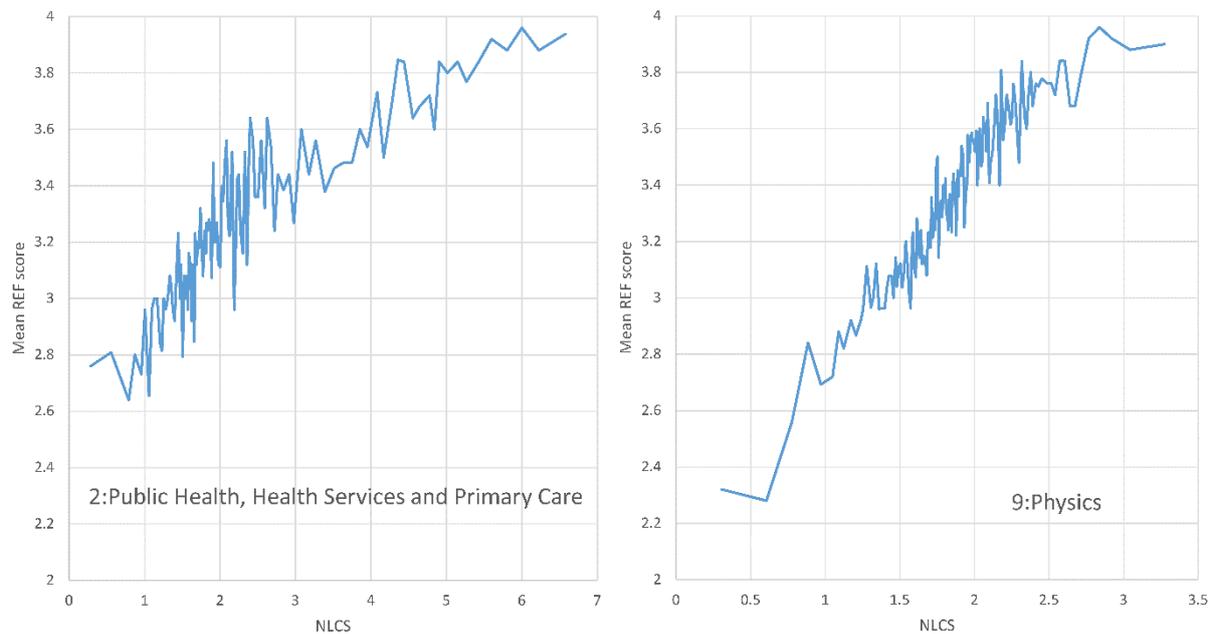

Figure 5. Mean REF scores for 2014-18 journal articles submitted to UK REF2021 in UoAs 2 and 9 against field and year normalised citation counts (NLCS). Articles are bucketed into groups of at least 25 with similar NLCS.

In UoAs where the correlation between NLCS and REF scores is more moderate, the slope of the broadly linear trend between NLCS and average REF2021 scores is less steep but still clear and does not get as close to the maximum (Figures 7, 8). In these fields, whilst there is a tendency for more cited articles to be higher quality, many articles break this trend.

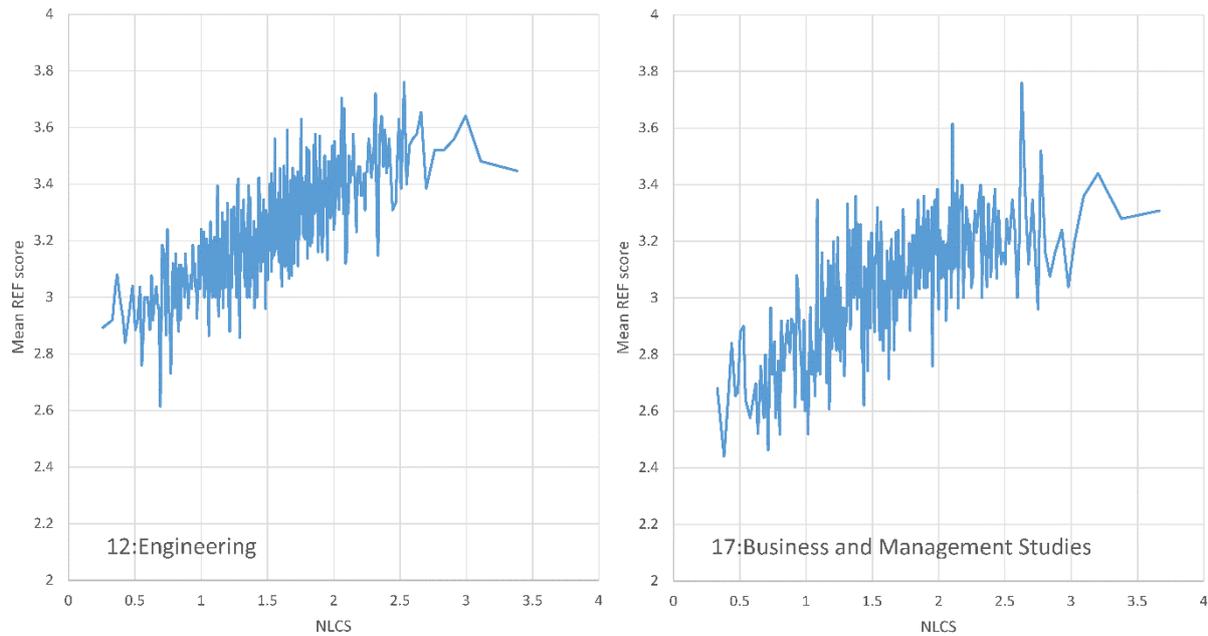

Figure 6. Mean REF scores for 2014-18 journal articles submitted to UK REF2021 in UoAs 12 and 17 against field and year normalised citation counts (NLCS). Articles are bucketed into groups of at least 25 with similar NLCS.

In UoAs where the correlation between REF2021 scores and NLCS is close to 0, this probably reflects a very shallow increasing tendency rather than a more complex relationship (e.g., not a U-shaped curve) (Figure 7). A shallow general slope like that for UoA 26 may reflect combinations of fields, some of which have no relationship between citations and quality (e.g., modern languages) and others that have some relationship (e.g., computational linguistics).

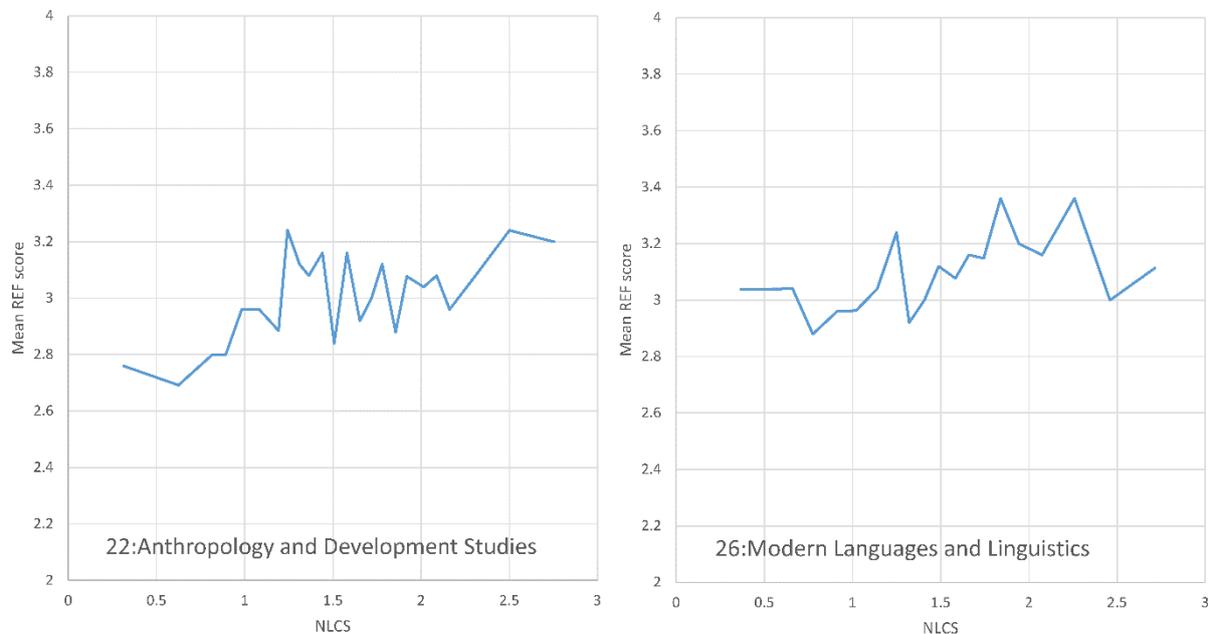

Figure 7. Mean REF scores for 2014-18 journal articles submitted to UK REF2021 in UoAs 22 and 26 against field and year normalised citation counts (NLCS). Articles are bucketed into groups of at least 25 with similar NLCS.

The shapes for the remaining 28 UoAs are broadly similar and all except two are in the supplementary file.

## 5 Discussion

Although this is the largest study of its kind it has many limitations. First, all journal articles are from the UK and the relationship between citations and quality (and its different operationalisations) might be different in other countries, such as those that value research applications more highly than scientific contributions. Second, the articles are self-selected and represent the outputs considered by the authors to be their best work. The relationship might be different for lower quality research. Third, the field normalisation is limited by the primarily journal-based categorisation scheme of Scopus, which might generate anomalies through interdisciplinary journals. For example, the large *Journal of the Acoustical Society of America* is classified as both Acoustics and Ultrasonics (Physics and Astronomy) and Arts and Humanities (misc.). Its relatively highly cited articles in the latter category (e.g., "Grid-free compressive beamforming" with 99 Scopus citations) will greatly add to the denominator of the field normalisation calculations, and reduce the field normalised scores of genuinely arts and humanities research in its category. Fourth, there may well be narrow fields (and other output types) for which the relationship between citations and research quality is inverted or null. Finally, the bucket size used to investigate RQ3 and RQ4 is relatively arbitrary.

The results update, extend and replace with a more rigorous academic study, the largest prior document-level related investigation (HEFCE, 2015), by showing for the first time with extra statistical power and field classification systems, that a positive relationship between research quality and citations is relatively universal. It was already known that the strength of the relationship varied between fields at the institutional level (e.g., Franceschet & Costantini, 2011; HEFCE, 2015, Mahdi et al., 2008) and suspected for articles (HEFCE, 2015), but not its universally positive nature. Although not all correlation confidence intervals excluded zero, the correlations were positive for all 34 UoAs, all 22 FOR codes and all 27 Scopus broad fields. Out of these, only three (very wide) confidence intervals contained 0 and these covered few articles (UoA 29 Classics [n=70] and UoA 31 Theology [n=124] in Figure 1 and Dentistry [n=115] in Figure 3). Thus, whilst not fully proven, the results are consistent with a positive relationship occurring across all broad academic fields, and give strong evidence that the relationship is near universal. The statistical power of the large numbers of articles in many fields supports this conclusion even for fields where the correlation is weak.

The unexpected finding of a positive (albeit weak) association between citations and quality across the arts and humanities has multiple plausible explanations. It is possible that all arts and humanities categories in all three schemes had a degree of pollution by social science or science articles, which was enough to create a detectable association. Alternatively, citations to articles may reflect influence (i.e., an aspect of quality) often enough in the arts and humanities to be detectable amongst the noise of other types of citation. Since the association is weak, a lot of empirical evidence, such as from citation motivation surveys, would be needed to distinguish between the two.

The finding that there is no reasonable citation threshold (field and year normalised) above which all articles are world leading research confirms that citation counts are never fully effective substitutes for human judgement, even in extreme cases. Whilst it is well known that articles occasionally become highly cited for negative reasons (e.g., the MMR/autism study: Godlee et al., 2011; the cold fusion article: Berlinguette et al., 2019), the results suggest that is it in fact *normal* for occasional articles in all fields to become extremely highly cited without having world leading quality. Moreover, in many fields (most UoAs) an extremely highly cited article is likely to be *not* world leading (e.g., averages below 3.5 in Figure 4). This does not undermine the use of percentiles in research evaluation, such as reporting the percentage of articles in the most cited 1% for a country (e.g., Rodriguez-Navarro, & Brito, 2022) but it cautions against fully equating highly cited with world leading research in any fields at the individual article level.

The close to linear relationship between the field and year normalised citation counts and research excellence is apparently the first finding of its kind. Whilst its primary value is the monotonically increasing nature of all graphs, showing the positive association occurs at all more most citation levels. It is also interesting as a theoretical issue but should not be interpreted at face value for two reasons, however. First, REF scores are ordinal rather than forming a scale: it is not clear that the gap between, say, 1* and 2* is the same as the gap between 3* and 4*, or even that the concept of gap width in this context is meaningful. In the absence of evidence to the contrary, it is reasonable to at least hypothesise that the scores form a numerical scale. Nevertheless, the citation counts are log transformed as part of the NLCS calculation, so the x axis of Figure 5 to 10 is effectively log-transformed. If the x-axes were reverse log transformed, expanding the difference between the higher numbers, then the graph shapes would be close to logarithmic. Thus, it is reasonable to hypothesise that the underlying relationship between research quality and citation counts is logarithmic, with citation counts providing diminishing returns in terms of increased probability of higher quality at higher values. This would fit with the rich-get-richer phenomenon by which highly cited articles are believed to attract new citations partly because they are highly cited rather than for their intrinsic value (Merton, 1968).

# 6 Conclusions

The universal positive association between citation scores and research quality should provide reassurance for those that appropriately use citation-based indicators to support research quality evaluations. They also suggest, unexpectedly and for the first time despite over half a century of citation analysis, that there are no broad fields of scholarship for which citations are *completely* irrelevant. Nevertheless, the wide variation between fields in the strength of the relationship confirms that citation-based indicators need greater levels of aggregation to yield useful information in some fields than others. For example, in fields with correlations above 0.5 at the article level, very strong aggregate correlations between average citations and average quality might be expected for small departments or small journals whereas the same aggregate correlations might only appear for very large departments or very large journals in other fields. Thus, the argument against inappropriate use of citations should not be that they are completely irrelevant in a field but that it is not reasonable to use them at a too low level of aggregation. Of course, if there are systematic biases in the citation data that field normalisation cannot eliminate, such as against qualitative research in a mixed methods field, then citation-based indictors would need to be used very cautiously in any context.

The fact that extremely high citation counts do not guarantee the highest research quality in any field and are not a high probability indicator of it in most (at least at the level of REF2021 UoAs) is another new finding. This should be remembered when journal articles are ranked by citations to identify the most influential articles in a field (Shadgan et al., 2010). For example, in June 2022 Google Scholar reported 302 articles containing the phrase "top cited articles" and 69 for "top cited papers" and many other bibliometric investigations include lists of top cited articles even if the investigations do not focus on highly cited papers. Moreover, some research evaluations count the top proportion of articles in the top 1% cited as an indicator of capacity to produce excellent research ("the vanguards of science": Wagner et al., 2022). In these contexts, it should always be recalled that articles can become highly cited for reasons other than research excellence.

**FUNDING INFORMATION**
This study was funded by Research England, Scottish Funding Council, Higher Education Funding Council for Wales, and Department for the Economy, Northern Ireland as part of the Future Research Assessment Programme (https://www.jisc.ac.uk/future-research-assessment-programme). The content is solely the responsibility of the authors and does not necessarily represent the official views of the funders.
**DATA AVAILABILITY**
Extended information about the data can be found in an associated report (http://cybermetrics.wlv.ac.uk/TechnologyAssistedResearchAssessment.html). The raw data was deleted before submission to follow UKRI data protection policy for REF2021.